\documentstyle[prl,aps,amsfonts,amssymb,epsf]{revtex}
\begin{document}
\draft
\twocolumn[\hsize\textwidth\columnwidth\hsize\csname@twocolumnfalse\endcsname

\title{
Evidence of first-order transition between vortex glass and Bragg glass phases\\
in high-$T_{\rm c}$ superconductors with point pins: Monte Carlo simulations
}

\author{Yoshihiko Nonomura\cite{nonomail} and Xiao Hu}
\address{
National Research Institute for Metals, Tsukuba, Ibaraki 305-0047, Japan}
\date{Received 14 February 2000}
\maketitle
\begin{abstract}
Phase transition between the vortex glass and the Bragg 
glass phases in high-$T_{\rm c}$ superconductors 
in $\vec{B}\parallel\vec{c}$ is studied by Monte 
Carlo simulations in the presence of point pins. 
A finite latent heat and a $\delta$-function peak of 
the specific heat are observed, which clearly indicates 
that this is a thermodynamic first-order phase transition. 
Values of the entropy jump and the Lindemann number 
are consistent with those of melting transitions. 
A large jump of the inter-layer phase difference is consistent 
with the recent Josephson plasma resonance experiment of 
Bi$_{2}$Sr$_{2}$CaCu$_{2}$O$_{8+y}$ by Gaifullin {\it et al.} 
\end{abstract}
]
%% \narrowtext
%
Vortex states in high-$T_{\rm c}$ superconductors have been 
intensively studied experimentally and theoretically \cite{Blatter}. 
Because of large fluctuations owing to high transition temperature 
and strong anisotropy, the flux-line lattice (FLL) melts at much 
lower temperatures than those predicted by Abrikosov's mean-field 
theory. The FLL melting is a thermodynamic first-order 
phase transition. In pure systems, the melting line stretches 
up to a high magnetic field as large as $H_{{\rm c}_{2}}$. 
However, all experiments show that first-order melting lines 
terminate at much lower magnetic fields \cite{Kwok,Zeldov}. 
Complicated phase diagrams are obtained experimentally in 
Bi$_{2}$Sr$_{2}$CaCu$_{2}$O$_{8+y}$ (BSCCO) \cite{Fuchs} 
and YBa$_{2}$Cu$_{3}$O$_{7-\delta}$ (YBCO) \cite{Nishizaki98}, 
and it is believed that effects of impurities are essential 
in real materials. For example, vacancies of oxygen atoms, 
which play the role of point pins to flux lines, cannot be 
excluded completely even in crystals of highest quality. 

\begin{figure}
\vspace*{-0.5cm}
\epsfxsize = 10cm
\epsffile{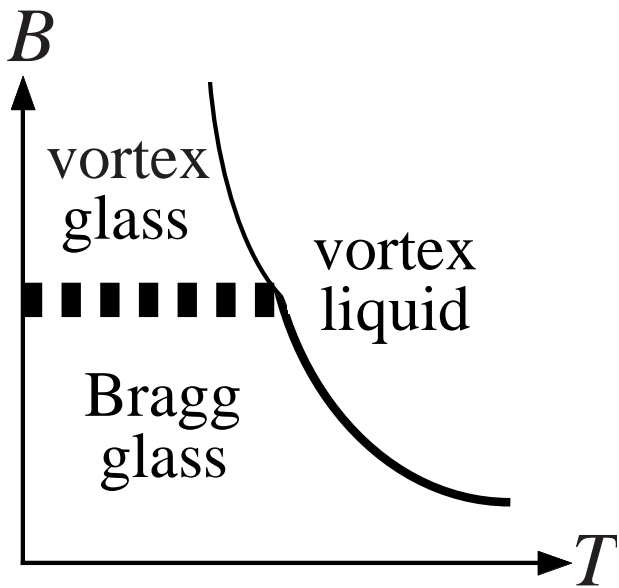}
\vspace{0.2cm}
\caption{
Schematic vortex phase diagram of high-$T_{\rm c}$ 
superconductors with point pins.
}
\label{phasepp}
\end{figure}
Fisher {\it et al.} \cite{Fisher} studied the Ginzburg-Landau 
model in a random potential, and proposed the so-called 
vortex glass (VG) phase before experimental studies. Giamarchi 
and Doussal \cite{Giamarchi} pointed out that the Bragg glass 
(BG) phase can exist in weak fields. In this phase, correlation 
functions decay in power laws \cite{Nattermann,Giamarchi94}, 
and the structure factor shows a triangular Bragg pattern. Accordingly, 
a phase transition between these two glass phases may be observed 
when the magnetic field is sweeped across the phase boundary 
(see Fig.\ \ref{phasepp}). The existence of the VG--BG phase 
transition in vortex systems with point pins shows a sharp 
contrast to the phase diagram of pure systems. Presuming 
a first-order phase transition, the VG--BG phase boundary 
was evaluated phenomenologically \cite{Ertas,Kierfeld}. 
However, physical properties around this phase boundary have 
not been clarified in experiments and numerical calculations 
until recently. The shape of the phase boundary seems to depend 
on observed quantities, and some experiments even suggest 
a crossover rather than a phase transition. Although some 
simulations \cite{Gingras,Ryu,Otterlo,Sugano} gave phase 
diagrams similar to that of Giamarchi and Doussal, numerical 
accuracy of these studies was not good enough to distinguish 
phase transitions and crossovers. The stability of the VG 
phase was studied by Kawamura \cite{Kawamura} 
including the screening effect. 

Quite recently, Gaifullin {\it et al.} observed \cite{Gaifullin} 
a large jump of the inter-layer phase difference on the VG--BG 
phase boundary of BSCCO by the Josephson plasma resonance 
experiment. They claimed that their observation is the evidence 
of a first-order phase transition. In the present Letter, we 
show more direct evidence of the first-order phase transition 
on the VG--BG boundary by large-scale Monte Carlo simulations. 
That is, a finite latent heat and a $\delta$-function peak of the 
specific heat are observed. Sharp jumps of the inter-layer phase 
difference and the averaged fluctuations of flux lines are also obtained. 

\begin{figure}
\epsfxsize = 8cm
\epsffile{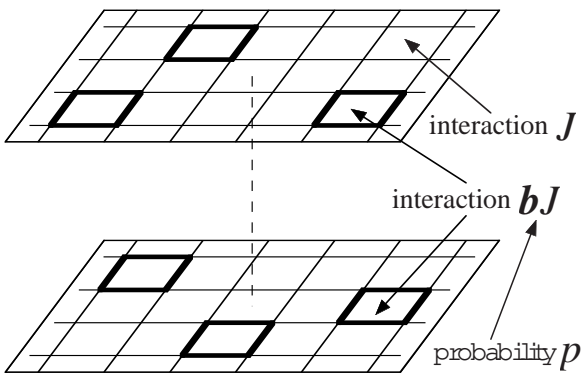}
\vspace{0.5cm}
\caption{Schematic description of point pins in the present model.}
\label{modelpp}
\end{figure}

In order to clarify vortex states and phase transitions of 
high-$T_{\rm c}$ superconductors in the presence of point pins, 
we start from the three-dimensional anisotropic, frustrated XY model 
on a simple cubic lattice \cite{Li,Hu}. Effects of point pins are 
introduced into the model by randomly-distributed weakly-coupled 
plaquettes in the $ab$ plane. Since a vortex sitting on a 
plaquette costs an energy proportional to the couplings 
surrounding it, flux lines tend to penetrate plaquettes 
with weaker couplings in order to reduce such loss of 
energies. The Hamiltonian of our model is given by 
\begin{eqnarray}
  \label{XYham}
  H&=&-\hspace{-0.2cm}
                 \sum_{i,j \in ab\ {\rm plane}} \hspace{-0.3cm}
                 J_{ij}\cos \left(\phi_{i}-\phi_{j}-A_{ij}\right)
                 \nonumber\\
           &&-\frac{J}{\Gamma^{2}} \hspace{-0.1cm}
               \sum_{m,n\parallel c\ {\rm axis}}\hspace{-0.2cm}
               \cos \left(\phi_{m}-\phi_{n}\right)\ ,\\
  A_{ij}&=&\frac{2\pi}{\Phi_0}\int^{j}_{i}{\bf A}^{(2)}
                                      \cdot {\rm d}{\bf r}^{(2)},
\end{eqnarray}
with the periodic boundary condition along all the directions. 
Couplings in the $ab$ plane are given by $J_{ij}=bJ$ ($0<b<1$) 
on the weakly-coupled plaquettes, and $J_{ij}=J$ otherwise. 
The density and the strength of point pins are controlled 
by the probability of weakly-coupled plaquettes, $p$, and 
the parameter $b$, respectively (see Fig.\ \ref{modelpp}). 
The pinning energy is of order of $(1-b)J$. A uniform magnetic 
field is applied along the $c$ axis, and its strength is proportional 
to the averaged number of flux lines per plaquette, $f$. 
Here we concentrate on the model with $L_{x}=L_{y}=50$ 
and $L_{c}=40$. This system size is large enough to describe 
the melting transition in the pure system ($b=1$) \cite{Hu}. 

In our model, we have four adjustable parameters: 
the anisotropy constant $\Gamma$, the density of flux lines $f$, 
the density of point pins $p$, and the strength of point pinning $b$. 
In order to investigate the VG--BG transition, we vary $b$, 
while fix the temperature at $T=0.06 J/k_{\rm B}$ and other 
parameters at $\Gamma=20$, $f=1/25$ and $p=0.003$. 
In other words, material parameters of the bulk system and the 
number and positions of point pins are not changed during 
the simulations. As will be shown later, this temperature is 
low enough for the study of the VG--BG phase boundary. 
Typical Monte Carlo steps (MCS) with the Metropolis algorithm 
are $3\sim 4\times 10^{7}$ MCS for equilibration, and 
$0.5\sim 1\times 10^{7}$ MCS for measurement. 

\begin{figure}
\vspace*{0.5cm}
\epsfxsize = 8.5cm
\epsffile{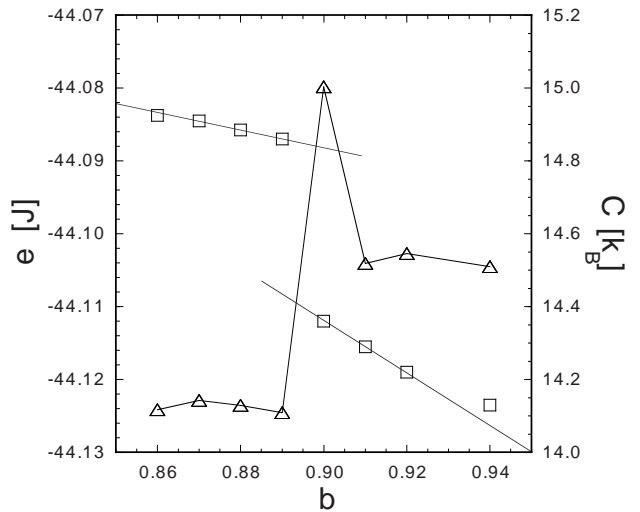}
\vspace{0.1cm}
\caption{
Internal energy $e$ (squares) and specific heat $C$ (triangles) 
per flux line per $ab$ plane versus strength of point pinning $b$. 
Straight lines are drawn as guides for eyes. 
}
\label{eC}
\end{figure}
First, the $b$ dependence of the internal energy $e$ and the 
specific heat $C$ per flux line per $ab$ plane is displayed in 
Fig.\ \ref{eC}. Clearly, the internal energy shows a sharp jump 
at the transition point $b^{\ast}=0.895\pm 0.005$ with a latent 
heat per flux line per $ab$ plane $Q\approx 2.3\times 10^{-2}J$, 
and a $\delta$-function peak of the specific heat also occurs at 
the same parameter. These two facts indicate that the VG--BG 
transition is a thermodynamic first-order phase transition. From 
this latent heat, the entropy jump at $b^{\ast}$ is estimated as 
\begin{equation}
  \Delta S=Q/T\approx 0.38 k_{\rm B}, 
\end{equation}
which is comparable to the experimental value in the melting transition 
of YBCO, $\Delta S\approx 0.5 k_{\rm B}$ \cite{Schilling}. 

\begin{figure}
\epsfxsize = 7.5cm
\epsffile{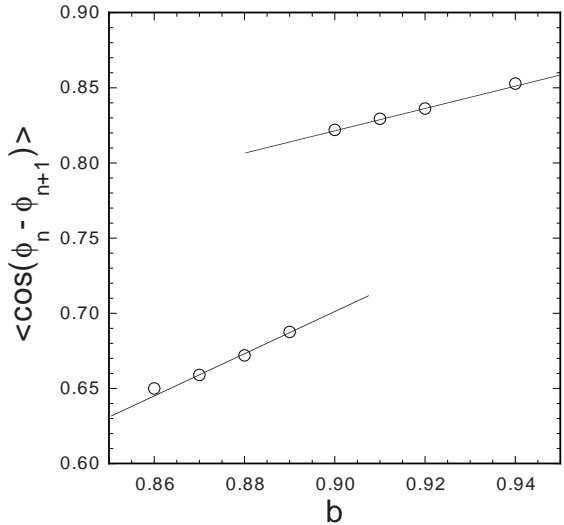}
\vspace{0.1cm}
\caption{
Inter-layer phase difference 
$\langle\cos(\phi_{n}-\phi_{n+1})\rangle$ versus strength 
of point pinning $b$. Straight lines are drawn as guides for eyes. 
}
\label{cosf}
\end{figure}

Second, the $b$ dependence of the inter-layer phase difference, 
$\langle\cos(\phi_{n}-\phi_{n+1})\rangle$, is plotted in 
Fig. \ref{cosf}. This quantity is related to the Josephson energy per 
phase variable $e_{\rm J}$ and the anisotropy constant $\Gamma$ as 
\begin{equation}
  \langle\cos(\phi_{n}-\phi_{n+1})\rangle=-e_{\rm J}\Gamma^{2}/J, 
\end{equation}
and a small change of $e_{\rm J}$ is magnified in this quantity in 
extremely anisotropic systems. This quantity also jumps sharply 
at $b^{\ast}=0.895\pm 0.005$, and the value of the jump at 
$b^{\ast}$, $\Delta_{\rm PD}\approx 0.12$, is as large as the 
experimental value, $\Delta_{\rm PD}\approx 0.2$ \cite{Gaifullin}. 
Moreover, the ratio of the jump of the Josephson energy to the 
latent heat is given by $\Delta e_{\rm J}/(Qf)\approx 0.34$, 
which means that the latent heat is equally distributed 
to all the directions in the VG--BG phase transition 
in extremely anisotropic systems.

\begin{figure}
\vspace*{0.5cm}
\epsfxsize = 7.5cm
\epsffile{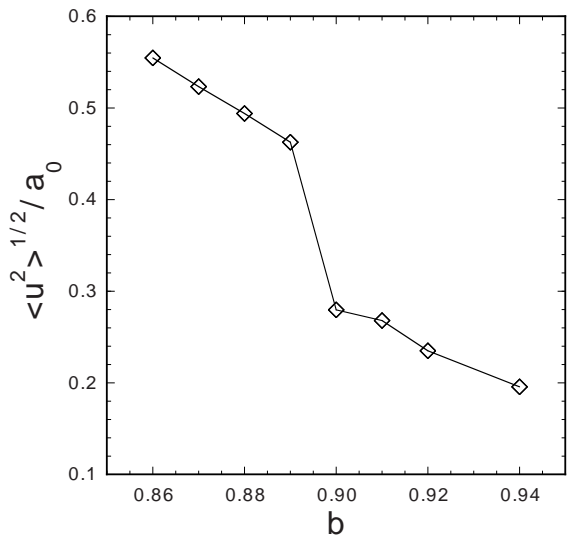}
\vspace{0.1cm}
\caption{Averaged fluctuations of flux lines 
$\langle u^{2}\rangle^{1/2}/a_{0}$ 
versus strength of point pinning $b$.}
\label{Lindemann}
\end{figure}

Third, the Lindemann number is evaluated directly \cite{Nonomura99}. 
The deviation $u$ of a flux line is measured in each $ab$ plane from 
the projection of the mass center of the flux line, and averaged over 
all the flux lines and the $ab$ planes. Then, the Lindemann number 
$c_{\rm L}$ is given by 
\begin{equation}
  c_{\rm L}=\lim_{b\to b^{\ast}+0}
            \langle u^{2}\rangle^{1/2}/a_{0}\ ,
\end{equation}
with the lattice constant of the triangular FLL, 
$a_{0}=(2/\sqrt{3})^{1/2}/f^{1/2}$. 
The $b$ dependence of $\langle u^{2}\rangle^{1/2}/a_{0}$ 
is shown in Fig.\ \ref{Lindemann}, and we have 
$c_{\rm L}\approx 0.28$. This value is almost equal to the one 
obtained in the FLL melting of pure systems \cite{Nonomura99}. 

Finally, we go into some details of the present simulations. 
The system with $b=0.90$ is calculated at first. Simulations are 
started from a random configuration at a very high temperature, 
and the system is gradually cooled down to $T=0.06 J/k_{\rm B}$. 
During the cooling process, the first-order melting transition 
characterized by a discontinuous appearance of the helicity 
modulus along the $c$ axis, $\Upsilon_{c}$ \cite{Hu}, 
takes place at $T_{\rm m}\approx 0.079 J/k_{\rm B}$, 
which corresponds to the vortex liquid (VL)--BG phase 
transition. Then, the strength of point pinning $b$ is varied. 
Since the quantity $\Upsilon_{c}$ is proportional to the 
superfluid density, the region with finite $\Upsilon_{c}$ 
is superconducting. This quantity is nonvanishing for all the 
values of $b$ shown in Figs.\ \ref{eC}--\ref{Lindemann} 
at $T=0.06 J/k_{\rm B}$, and therefore the phase transition 
investigated in the present Letter is not the VL--BG one, but 
the VG--BG one. Equilibration in systems with point pins is 
much slower than that in pure systems, and only one sample 
can be taken for calculations at present. Nevertheless, the results 
obtained in the present Letter are quite clear-cut and consistent with 
experiments. Thus, the small number of random sampling does 
not seem serious. Since positions of point pins are independent 
in each $ab$ plane, the number of $ab$ planes, $L_{c}=40$, 
would be large enough for averaging effects of point pins. 

Although we have concentrated on the VG--BG transition for a 
single density of point pins $p$ in the present Letter, we have also 
investigated the VL--BG and VL--VG transitions for various $p$, 
and obtained the overall phase diagram in the $p$--$T$ plane. 
The structure of the $p$--$T$ phase diagram is similar to 
that of the $B$--$T$ phase diagram. Experimentally, the 
increase of $p$ corresponds to the repeated irradiation of 
electrons or protons, and our $p$--$T$ phase diagram is 
consistent with recent experiments \cite{Nishizaki00,Paulius}. 
Details of this study will be reported elsewhere \cite{Nonomura00}. 

In conclusion, the first thermodynamic evidence of the first-order 
transition between the vortex glass (VG) and the Bragg glass (BG) 
phases has been obtained in high-$T_{\rm c}$ superconductors in 
the presence of point pins. A finite latent heat and a $\delta$-function 
peak of the specific heat are observed by large-scale Monte Carlo 
simulations of the three-dimensional anisotropic, frustrated XY 
model with randomly-distributed weakly-coupled plaquettes. 
The entropy jump derived from the latent heat is nearly 
equal to those in the melting transition of YBCO. 
The Lindemann number evaluated from fluctuations of 
flux lines, $c_{\rm L}\approx 0.28$, is reasonable for the 
first-order phase transition. The inter-layer phase difference 
also shows a sharp jump on the VG--BG phase boundary. 
This property is consistent with the Josephson plasma 
resonance experiment of BSCCO by Gaifullin {\it et al.}

The present authors would like to thank Prof.~Y.~Matsuda 
for communications. Numerical calculations were performed 
on Numerical Materials Simulator (NEC SX-4) at National 
Research Institute for Metals, Japan. 

\narrowtext
\end{document}